\documentclass[fleqn,useAMS,usenatbib]{mnras}
\usepackage[english]{babel}
\usepackage{fancyhdr}
\usepackage{amsmath}
\usepackage{amssymb}
\usepackage{multicol}
\usepackage{layout}
\usepackage{graphicx}
\usepackage{multirow}
\usepackage{xcolor}
\usepackage{txfonts}
\usepackage[T1]{fontenc}
\usepackage{times}
\usepackage{natbib}

\newif\ifAMStwofonts
\AMStwofontstrue

\graphicspath{{./fig/}}
\usepackage{aecompl}

\title[Shear and rotation in massive galaxy clusters]{Constraints on shear and rotation with massive galaxy clusters}

\author[Mehrabi et~al.]{Ahmad Mehrabi$^{1}$ \thanks{mehrabi@basu.ac.ir}, Francesco Pace$^2$, Mohammad Malekjani$^{1}$ 
and Antonino Del Popolo$^{3,4,5}$ \\
$^1$ Department of Physics, Bu-Ali Sina University, Hamedan 65178, Iran\\
$^2$ Jodrell Bank Centre for Astrophysics, School of Physics and Astronomy, The University of Manchester,
Manchester, M13 9PL, U.K.\\
$^3$ Dipartimento di Fisica e Astronomia, Universit\`a di Catania, Viale Andrea Doria 6, 95125 Catania, Italy \\
$^4$ INFN sezione di Catania, Via S. Sofia 64, I-95123 Catania, Italy \\
$^5$ International Institute of Physics, Universidade Federal do Rio Grande do Norte, Brazil
}

\date{Accepted ?, Received ?; in original form \today}

\pagerange{\pageref{firstpage}--\pageref{lastpage}} \pubyear{2016}

\begin{document}

\label{firstpage}

\maketitle

\begin{abstract}
A precise determination of the mass function is an important tool to verify cosmological predictions of the 
$\Lambda$CDM model and to infer more precisely the better model describing the evolution of the Universe. Galaxy 
clusters have been currently used to infer cosmological parameters, in particular the matter density parameter 
$\Omega_{\rm m}$, the matter power spectrum normalization $\sigma_8$ and the equation of state parameter $w_{\rm de}$ 
of the dark energy fluid. 
In this work, using data on massive galaxy clusters ($M>8\times 10^{14}~h^{-1}~M_{\odot}$) in the redshift range 
$0.05\lesssim z\lesssim 0.83$, for the first time we put constraints on the parameter $\alpha$ introduced within 
the formalism of the extended spherical collapse model to quantify deviations from sphericity due to shear and 
rotation. 
Since at the moment there is no physical model describing its functional shape, we assume it to be a logarithmic 
function of the cluster mass. By holding $\sigma_8$ fixed and restricting our analysis to a $\Lambda$CDM model, we 
find, at $1-\sigma$ confidence level, $\Omega_{\rm m}=0.284\pm0.0064$, $h=0.678\pm0.017$ and 
$\beta=0.0019^{+0.0008}_{-0.0015}$, where $\beta$ represents the slope of the parameter $\alpha$. This results 
translates into a $9\%$ decrement of the number of massive clusters with respect to a standard $\Lambda$CDM mass 
function, but better data are required to better constrain this quantity, since at the $2-\sigma$ and $3-\sigma$ 
confidence level we are only able to infer upper limits.
\end{abstract}

\begin{keywords}
 cosmology: methods: analytical - statistical - cosmology: theory - dark energy
\end{keywords}

\section{Introduction}

It is strongly believed that large scale cosmic structures such as galaxies and clusters of galaxies are originated 
from small initial fluctuations during the inflationary era \citep{Starobinsky1980,Guth1981,Linde1990}. 
Later, these fluctuations can grow due to gravitational instability 
\citep{Gunn1972,Press1974,White1978,Peebles1993,DelPopolo1998,Peacock1999,Sheth1999,Barkana2001,Peebles2003,Ciardi2005,
Bromm2011}. 
Most of the growth of cosmic structures takes place after the decoupling epoch between photons and baryons. At early 
times, when the amplitude of fluctuations is very small, linear perturbation theory can be safely used to study the 
evolution of fluctuations. However, at later times, when the amplitude of fluctuations becomes large, linear 
perturbations theory fails because fluctuations enter in the non-linear regime. Hence, more sophisticated techniques 
are required. 
The spherical collapse model (SCM), first introduced by \cite{Gunn1972}, is a simple analytical method to follow the 
non-linear evolution of the growth of fluctuations on sub-Horizon scales. This model has been widely investigated in 
literature 
\citep{Fillmore1984,Bertschinger1985,Hoffman1985,Ryden1987,Subramanian2000,Ascasibar2004,Williams2004,Mota2004,
Maor2005,Basilakos2009,DelPopolo2009,Li2009,Pace2010,Wintergerst2010,Pace2012,Pace2014b,Naderi2015,Malekjani:2015pza}. 
Moreover, in the SCM, the wavelengths of perturbations are much smaller than the Hubble radius and therefore the
Pseudo-Newtonian (PN) hydrodynamical equations can be applied in this formalism \citep{Lima:1996at}. 
It has been shown that the results of the quoted approach in the linear regime are consistent with general relativity 
(GR) theory \citep{Abramo2007,Abramo:2008ip}. Recently, the SCM has been extended to more general cases where also the 
rotation (vorticity) $\omega$, and the shear $\sigma$, are taken into account 
\citep{Popolo2013a,Popolo2013b,Popolo2013c,Pace2014a}. \cite{Reischke2016} instead provided a description of the shear 
due to tidal shear forces using the Zel'dovich approximation, hence not relying on phenomenological approaches. 
In particular, \cite{Popolo2013b,Popolo2013c} showed that in the presence of shear and rotation, the collapse of 
structures is slowed down due to the strength of the rotation term, the SCM parameters become mass dependent as in the 
ellipsoidal collapse (with a stronger dependence on galactic scales).

One of the most important features of the SCM formalism is that it can be used to describe the abundance of collapsed 
haloes as a function of mass and redshift \citep{Press1974}. Information coming from the abundance of collapsed 
structures is an important tool to study the distribution of matter in the universe \citep{Evrard:2001hu}. 
Observationally, the mass function and the number counts of massive galaxy clusters have been inferred through X-ray 
surveys \citep{1999A&A...344...17D,Borgani:2001ir,Reiprich:2001zv,2009ApJ...692.1033V}, weak and strong lensing studies 
\citep{Bartelmann:1997xa,Dahle:2006fa,Corless:2009hi,Corless:2008wk} and optical surveys, like the SDSS 
\citep{Bahcall:2003sq,Bahcall:2002wx,Bahcall:2003hu}. 
It should be noted that the redshift evolution of massive clusters depends strongly on cosmological parameters, 
especially on the amplitude of mass fluctuations $\sigma_8$ and on the non-relativistic matter density parameter 
$\Omega_{\rm m}$ \citep{Bahcall:1998ur,Bahcall:2002ru}. Higher values of $\sigma_8$ favour the formation of haloes at 
early times, while lower values give rise to fewer massive clusters at high redshifts. 
In the last two decades, data on the number counts of massive clusters 
($M_{\rm cluster}>8\times10^{14}h^{-1}M_{\odot}$ in a comoving radius $R_{\rm cluster}=1.5h^{-1}$~Mpc) at low and high 
redshifts ($0.05\leq z\leq 0.83$) were used to determine the linear amplitude of mass fluctuations and the 
non-relativistic matter density in a Universe with a cosmological constant \citep{Bahcall:1998ur,Bahcall:2002ru}. 
Recently, these data have been used to put constraints on some of the free parameters of the standard cosmological 
model and to investigate the possibility that dark energy evolves in time, instead of being constant 
\citep{Campanelli2012}. 
Also, \cite{Devi:2011gb} constrained different dark energy models by using the number count data of massive clusters 
in the context of the SCM. Their results show that the cluster number density for different dark energy models 
significantly deviates from that obtained in the concordance $\Lambda$CDM Universe, especially at high redshifts. 
Furthermore, in scalar field DE models, \cite{Devi:2011gb} showed that the tachyon scalar field with a linear 
potential has the largest deviations from the $\Lambda$CDM model.

In this work, in the context of the extended SCM in the presence of shear and rotation (hereafter, ESCM), using the 
number count data of massive galaxy clusters, we put constraints on the parameter $\alpha$ related to the combined 
shear and rotation parameter in the equations dealing with the ESCM. To do this, we use the available number count data 
from massive X-ray clusters presented in \cite{Campanelli2012}. Since we want to focus on studying the contribution 
of the shear and rotation term in massive clusters, we limit ourselves to a standard $\Lambda$CDM cosmology. 
Moreover, we adopt the Navarro-Frenk-White (NFW) profile for the virialized halo mass density as found in N-body 
simulations \citep{Navarro1997}.

The paper is organised as following. In section~\ref{sect:2}, we present the ESCM and study the spherical collapse 
parameters in the presence of shear and rotation. In section~\ref{sect:3}, we describe how to evaluate the number 
counts of massive clusters and in section~\ref{sect:4} we constrain the cosmological parameters including the shear 
and rotation parameter by applying a Markov Chain Monte Carlo (MCMC) analysis, using SnIa, BAO, CMB, the Hubble 
parameter, the Big Bang Nucleosynthesis (BBN) and the number count data of massive clusters. 
Our results and conclusion are presented in section~\ref{sect:5}.

\section{Extended spherical collapse model}\label{sect:2}
In this section, we review the derivation of the differential equations determining the evolution of matter overdensity 
$\delta$, using the spherical collapse model in the presence of shear and rotation. The spherical collapse model in 
dark energy cosmologies was investigated in detail in \cite{Pace2010}, based on the work of \cite{Abramo2007}. 
\cite{Pace2010} extended the evolution equation to general geometries and cosmologies, so that their results may be 
applied to models beyond the $\Lambda$CDM model. The effects of shear and rotation on the evolution of matter 
overdensities were investigated in homogeneous DE cosmologies \citep{Popolo2013a,Popolo2013b,Popolo2013c} and in 
clustering DE cosmologies \citep{Pace2014a}. Using the non-linear differential equations for the evolution of the 
matter density contrast derived from Newtonian hydrodynamics in \cite{Pace2010}, \cite{Popolo2013b} showed that the 
parameters of the spherical collapse model become mass dependent. Due to the stronger effect of rotation with respect 
to shear, the linear overdensity parameter $\delta_{\rm c}$ is enhanced with respect to the standard case, therefore 
the collapse is slowed down and less objects will form in general. 
A similar effect was also obtained for the virial overdensity parameter $\Delta_{\rm V}$ \citep{Popolo2013b}. On the 
other hand, in the high mass tail of the mass function, they showed that the effects of shear and rotation are very 
small, not influencing in an appreciable way the number of objects at high mass.

The equations describing the evolution of the density contrast $\delta_j\equiv\delta\rho_j/\bar{\rho}_j$ and of the 
peculiar velocity $\vec{u}_j$, together with the Poisson equation for the peculiar potential $\phi$ are 
\citep{Pace2010,Batista2013,Pace2014a},
\begin{align}
 \dot{\delta}_{\rm j} + 3H(c_{{\rm eff},j}^2 - \bar{w}_{\rm j})\delta_{\rm j} +
 [1+\bar{w}_{\rm j}+(1+c_{{\rm eff},j}^2)\delta_{j}]\vec{\nabla}\cdot\vec{u}_{\rm j} & = 0\;,\label{eq:per1} \\
 \dot{\vec{u}}_{\rm j} + 2H\vec{u}_{\rm j} + (\vec{u}_{\rm j}\cdot\vec{\nabla}\vec{u}_{\rm j}) + 
 \frac{1}{a^2}\vec{\nabla}\phi & = 0 \;, \label{eq:per2} \\
 \nabla^2\phi - 4\pi G a^2\sum_{\rm k}\rho_{\rm k}\delta_{\rm k}(1+3c_{{\rm eff},k}^2) & = 0 \;, \label{eq:enis0}
\end{align}
where we already took into account the top-hat density profile for the density perturbation ($\vec{\nabla}\delta_j=0$) 
and we worked in comoving coordinates ($\vec{x}$), $\vec{\nabla}\equiv\vec{\nabla}_{\vec{x}}$. 
In the previous set of equations, $\bar{w}_{\rm j}=\bar{P}_{\rm j}/(\bar{\rho}_{\rm j}c^2)$ is the background 
equation-of-state parameter of the fluid and $c_{\rm eff,j}^2=\delta P_{\rm j}/(c^2\delta\rho_{\rm j})$ is the 
effective sound speed of perturbations in units of the speed of light.

Shear and rotation enter in the picture by taking the divergence of equation~\ref{eq:per2}, since we are interested in
the evolution of the divergence of the peculiar velocity $\theta\equiv\vec{\nabla}\cdot\vec{u}$
\begin{equation}
 \dot{\theta}+2H\theta+\frac{1}{3}\theta^2+\sigma^2-\omega^2+\frac{1}{a^2}\vec{\nabla}^2\phi=0 \label{eqn:fullEuler}\;,
\end{equation}
where $\sigma^2=\sigma_{ij}\sigma^{ij}$ is the shear tensor and $\omega^2=\omega_{ij}\omega^{ij}$ the rotation tensor
and are defined as
\begin{align}
 \sigma_{ij} & = \frac{1}{2}\left(\frac{\partial u^j}{\partial x^i}+\frac{\partial u^i}{\partial x^j}\right)-
                 \frac{1}{3}\theta\delta_{ij}\;,\\
 \omega_{ij} & = \frac{1}{2}\left(\frac{\partial u^j}{\partial x^i}-\frac{\partial u^i}{\partial x^j}\right)\;,
\end{align}
To obtain equation~\ref{eqn:fullEuler}, we used the following relation
\begin{equation}
 \nabla\cdot[(\vec{u}\cdot\nabla)\vec{u}] = \frac{1}{3}\theta^2 + \sigma^2 - \omega^2\;. \label{eq:decomposition}
\end{equation}

Following \cite{Fosalba:1997tn,Fosalba:1998da,Gaztanaga:1997nz,Engineer:1998um}, it is possible to relate the evolution 
of $\delta$ to the evolution of the radius $R$ of the overdensity
\begin{equation}
 \frac{d^2R}{dt^2} = \frac{4}{3}\pi G\rho R-(\sigma^2-\omega^2)\frac{R}{3}+\frac{\Lambda}{3}R
                   = -\frac{GM}{R^2}-(\sigma^2-\omega^2)\frac{R}{3}+\frac{\Lambda}{3}R\;,
\end{equation}
similar to the equation of the spherical collapse model taking into account the angular momentum
\citep{Peebles1993,Nusser:2000xn,Zukin:2010sp,Zukin:2010gx}
\begin{equation}
 \frac{d^2R}{dt^2} = -\frac{GM}{R^2}+\frac{L^2}{M^2 R^3}+\frac{\Lambda}{3}R
                   = -\frac{GM}{R^2}+\frac{4}{25}\Omega^2 R+\frac{\Lambda}{3}R\;,
 \label{eqn:spher}
\end{equation}
where we used the momentum of inertia of a sphere, $I=2/5 M R^2$. The last equation shows a close connection between
the vorticity $\omega$ and the angular velocity $\Omega$.

Considering the ratio of the rotational and gravitational terms, we get
\begin{equation}\label{alpha}
 \alpha=\frac{L^2}{M^3 R G}\;.
\end{equation}

Following the same argument for the rotation we obtain
\begin{equation}\label{eq:sigmm}
 \frac{\sigma^2-\omega^2}{H^2}=-\frac{3}{2}\alpha \sum_k \Omega_{k,0}g_k(a)(1+3c_{{\rm eff},k}^2) \delta_k\;,
\end{equation}
where $g_k(a)$ represents the time evolution of the fluid $k$ and we assumed that all the fluids are affected by 
shear and rotation in the same way.

The effects of shear and rotation on $\delta$ can be obtained by solving equations~\ref{eq:per1}, \ref{eq:per2} and 
\ref{eq:enis0}, together with equation~\ref{eq:sigmm}.

Specialising to the standard $\Lambda$CDM model where only matter is clustering ($\bar{w}_{\rm m}=c^2_{\rm eff,~m}=0$), 
the equation of motion for the overdensity $\delta_{\rm m}$ becomes
\begin{equation}
 \delta^{\prime\prime}_{\rm m}+\left(\frac{3}{a}+\frac{E^{\prime}}{E}\right)\delta^{\prime}_{\rm m}-
 \frac{4}{3}\frac{{\delta^{\prime}}^2_{\rm m}}{1+\delta_{\rm m}}-
 \frac{3}{2a^5E^2}(1-\alpha)\Omega_{\rm m,0}\delta_{\rm m}(1+\delta_{\rm m})=0 \;,\label{eqc1}
\end{equation}
where $E$ represents the time evolution of the Hubble function, $H(a)=H_0E(a)$ and is given by
\begin{equation}
 E^2(a)=\frac{\Omega_{\rm m,0}}{a^3}+\Omega_{\Lambda}\;.
\end{equation}

\begin{figure*}
 \centering
 \includegraphics[width=0.45\textwidth]{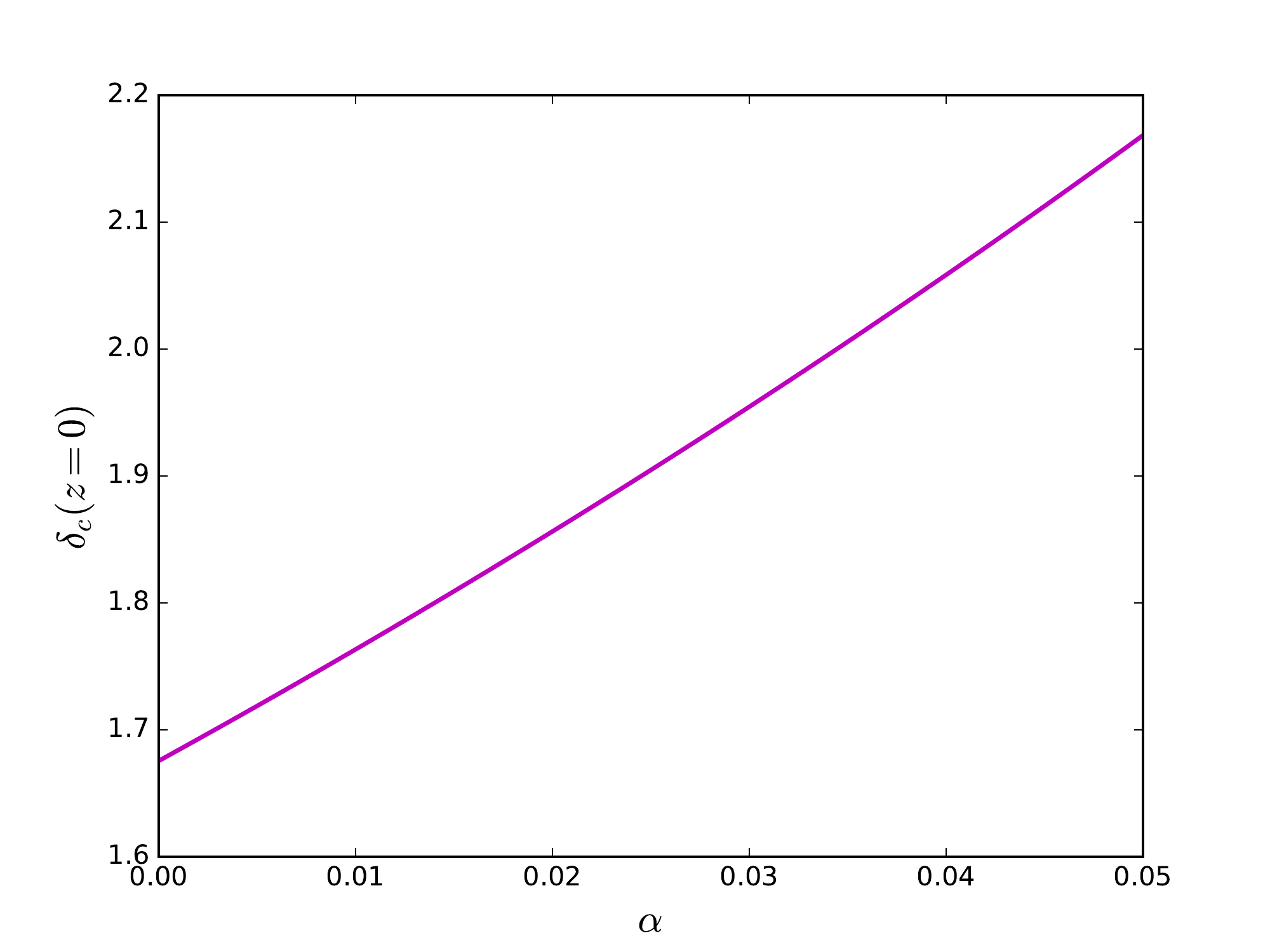}
 \includegraphics[width=0.45\textwidth]{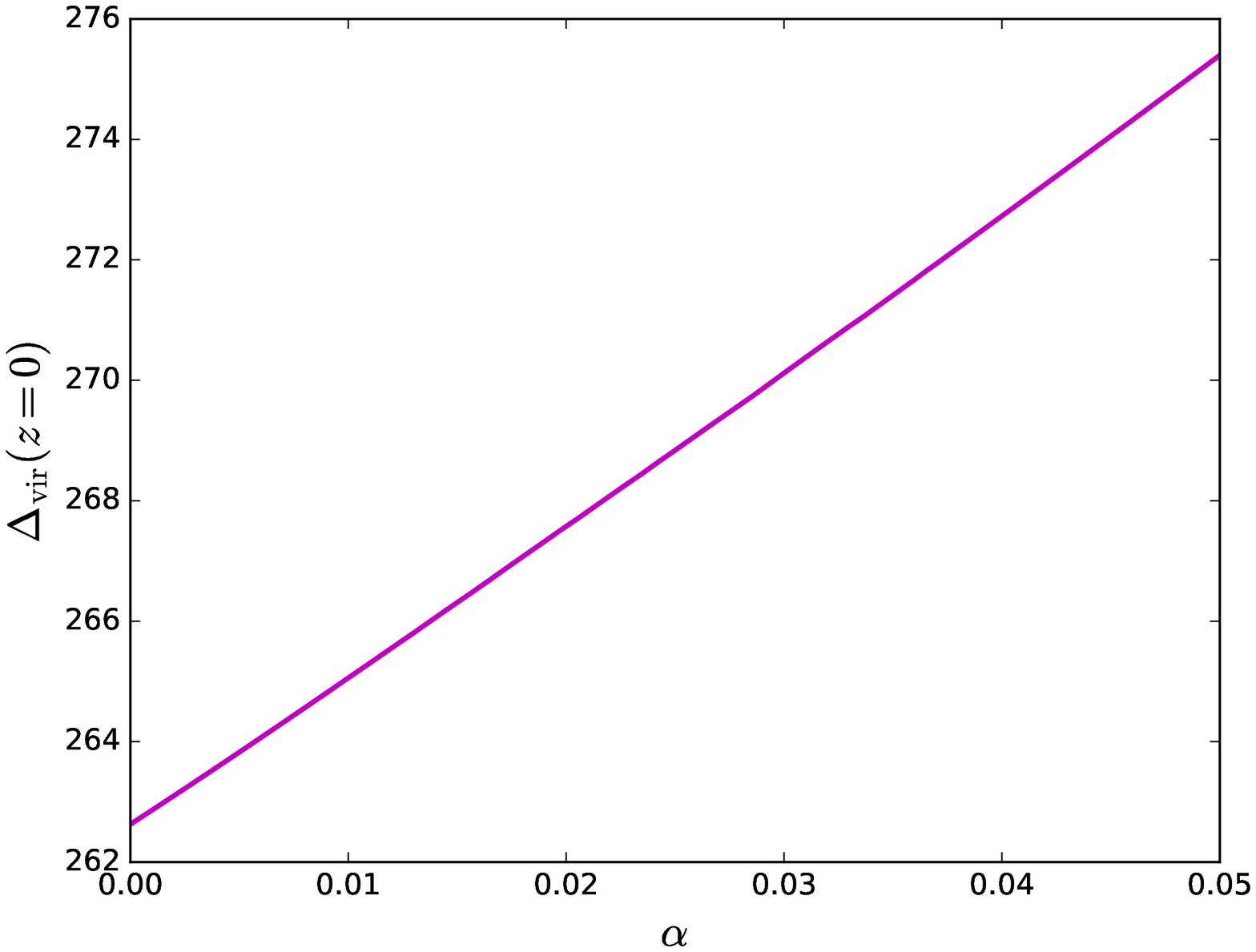}
 \caption{\textit{Left} (\textit{right}) \textit{panel}: Variation of the linear overdensity (virial overdensity 
 with respect to the background density) at the present time $\delta_{\rm c}(z=0)$ ($\Delta_{\rm vir}(z=0)$) in terms 
 of the parameter $\alpha$ in a $\Lambda$CDM universe.}
 \label{fig:delta_c}
\end{figure*}

Shear and rotation are non-linear quantities and will not affect the evolution of perturbations at the linear level.
Therefore the evolution of matter perturbations in the linear regime is the standard growth factor equation
\begin{equation}
 \delta^{\prime\prime}_{\rm m}+\left(\frac{3}{a}+\frac{E^{\prime}}{E}\right)\delta^{\prime}_{\rm m}-
 \frac{3}{2a^5E^2}\Omega_{\rm m,0}\delta_{\rm m}=0 \;.\label{eqc2}
\end{equation}

\subsection{Determination of \texorpdfstring{$\delta_{\rm c}$}{dc} and \texorpdfstring{$\Delta_{\rm vir}$}{DV} in the 
ESCM}
It is well known that the linear overdensity $\delta_{\rm c}$ and the virial overdensity $\Delta_{\rm vir}$ are two 
important quantities characterising the scenario of the SCM. Moreover, the first quantity is crucial in the 
Press-Schechter formalism \citep{Press1974,Bond1991,Sheth2002} and the latter determines the size of virialised 
structures. Here we calculate these two quantities in the context of the ESCM and study how shear and rotation can 
affect them. 
To this end, we follow the general approach presented in \cite{Pace2010,Pace2012,Pace2014a,Pace2014b} to calculate the 
linear overdensity $\delta_{\rm c}$ and virial overdensity $\Delta_{\rm vir}$ and we refer to these works for a more
detailed description.

Since equation~(\ref{eqc1}) is a non-linear equation, the value of $\delta_{\rm m}$ diverges at some characteristic 
redshift called the collapse redshift $z_{\rm c}$. Numerically, the divergence is achieved when $\delta_{\rm m}$ 
exceeds the value $10^7$ (this value is the minimum one necessary to have a solution numerically stable and independent 
of this value). The linear overdensity $\delta_{\rm c}$ is the value of the overdensity at collapse redshift 
$\delta_{\rm m}(z_{\rm c})$ obtained by solving the linearised equation (\ref{eqc2}) with the same initial conditions 
applied to the non-linear equation~(\ref{eqc1}).

In the left panel of Fig.~(\ref{fig:delta_c}), we plot the linear overdensity parameter at the present time, 
$\delta_{\rm c}(z=0)$, as a function of the parameter $\alpha$. In the case of $\alpha=0$, the linear overdensity 
$\delta_{\rm c}$ is $\approx 1.675$, as expected in a $\Lambda$CDM cosmology with $\Omega_{\rm m,0}=0.3$. 
By increasing $\alpha$, the value of $\delta_{\rm c}$ increases accordingly. Since in the presence of shear and 
rotation the collapse is delayed by their mutual interplay, we need a higher value for the initial overdensity to 
reach the collapse and this translates into a higher value of $\delta_{\rm c}$.

Another important quantity in the SCM is the virial overdensity, defined as the overdensity with respect to the 
background (or critical) density at the time of virialization. Note that this aspect is not native into the formalism 
and has to be introduced in it. This quantity is defined as $\Delta_{\rm vir}=\zeta (x/y)^3$, where $\zeta$ is the 
overdensity at the turn-around redshift, $x$ is the scale factor divided by the turn-around scale factor and $y$ is the 
ratio between the virialised radius and the turn-around radius \citep{Wang1998,Wang:2005ad}. 
In an Einstein-de Sitter (EdS) Universe, it is easy to show that $y=1/2$, $\zeta\simeq5.5$ and 
$\Delta_{\rm vir}\simeq 178$ at any time. However, in a $\Lambda$CDM universe, $\Delta_{\rm vir}$ is affected by the 
presence of the cosmological constant and the virial overdensity changes as a function of time.

In the right panel of Fig.~(\ref{fig:delta_c}), we show the evolution of the virial overdensity with respect to the 
background density computed at the present time, $\Delta_{\rm vir}(z=0)$, as a function of $\alpha$ for a $\Lambda$CDM 
cosmology within the framework of the ESCM. Dark energy opposes to gravity and prevents the collapse of structures. 
In the EdS model, this value is constant, but for the $\Lambda$CDM model and dark energy models in general, this is not 
the case any more. In particular, for a $\Lambda$CDM cosmology the value of $\Delta_{\rm vir}$ in standard SCM is 
roughly $260$ times the background density. In addition, in the ESCM due to the inclusion of the shear and rotation 
term, we see that $\Delta_{\rm vir}$ increases by increasing the parameter $\alpha$. This result is similar to what 
found for the linear overdensity $\delta_{\rm c}$ in the ESCM. 
We note that the SCM parameters $\delta_{\rm c}$ and $\Delta_{\rm vir}$ become mass dependent in the presence of 
shear and rotation as indicated in \cite{Popolo2013a,Popolo2013b,Popolo2013c}. In fact the effect of the parameter 
$\alpha$ on the SCM parameters can be easily seen from equation~(\ref{eqc1}). Since the non-linear quantity $\alpha$ 
depends on mass, we expect that in general the SCM parameters are mass-dependent and change for different mass scales. 
In the next section, we provide a phenomenological expression representing the variation of $\alpha$ in terms of the 
mass and show how the mass function and the number counts of massive galaxy clusters depend on $\alpha$.

 Since one of the main ingredients in our analysis is the virial overdensity $\Delta_{\rm vir}$, it is important to 
discuss a bit more in detail this issue. While it is straightforward to evaluate the virial overdensity in an EdS model 
where only the dark matter component is present, this is not true for more complicated models where also an additional 
fluid, either in the form of a cosmological constant $\Lambda$ or a more general dark energy component, is present. The 
main problem to face is the role of this additional component to the virialization process, not to mention further 
complications when the dark energy component clusters as well. Several authors have proposed different recipes to take 
this into account and results could differ sensibly between each other. 
\cite{Wang1998,Lokas2001,Basilakos2003,Horellou2005,Wang:2005ad,Basilakos2007} have studied the virialization process 
on smooth dark energy models and \citet{Bartelmann2006} have extended the formalism to early dark energy models 
\citep[but see also][]{Pace2010}; \cite{Maor2005} and \cite{Nunes2006} instead investigated the virialization process 
when dark energy possesses fluctuations. Finally, \cite{Basilakos2010} evaluated the mass function of cluster-size 
halos and their redshift distribution for different cosmological models and confronted to the predictions of the 
concordance $\Lambda$CDM model finding that the predictions of eight models were statistically different from the ones 
of the $\Lambda$CDM model. \\
This discussion clearly shows that a general consensus on the subject is still lacking and more importantly, that our 
discussion could be severely affected by the choice of the prescription to evaluate the virial overdensity and hence 
the size of cluster. We therefore decided to use the prescription to evaluate $\Delta_{\rm vir}$ of \cite{Wang1998} and 
\cite{Maor2005} which are in agreement with a fully general relativistically analysis performed by \cite{Meyer2012}.

\section{Massive galaxy cluster number count}\label{sect:3}
In this section we investigate the mass function and the number counts of massive galaxy clusters in the framework of 
the ESCM formalism and study how shear and rotation affect observable quantities related to galaxy clusters number 
counts.

\begin{figure}
 \centering
 \includegraphics[width=0.5\textwidth]{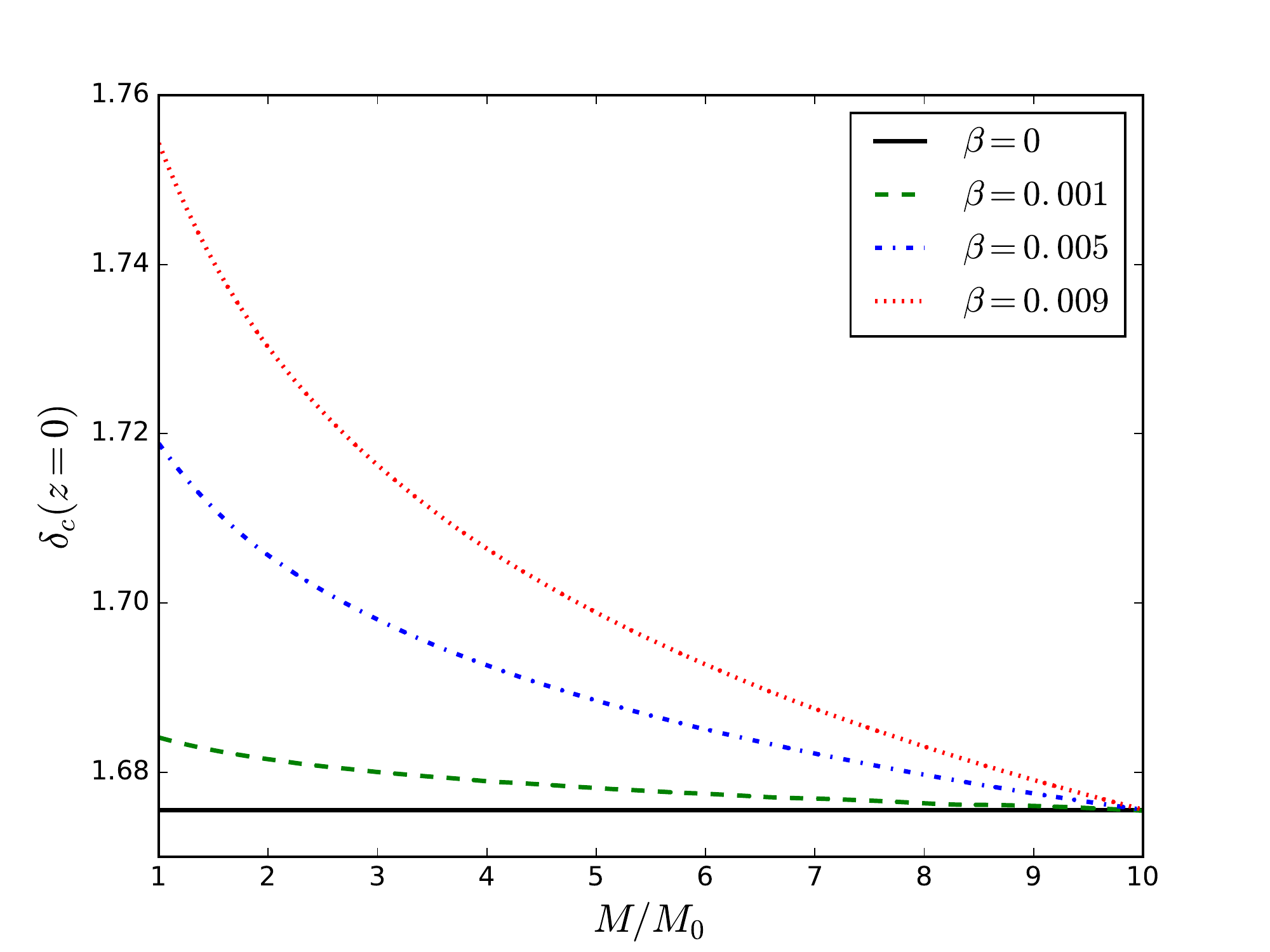}
 \caption{The present time value of the linear overdensity $\delta_{\rm c}(z=0)$ as a function of mass computed for 
 different values of the slope parameter $\beta$. The black solid line shows results for $\beta=0$ (the standard 
 spherically symmetric case), the green dashed curve the model with $\beta=0.001$, while the blue dot-dashed and the 
 red dotted line show results for $\beta=0.005$ and $\beta=0.009$, respectively.  The normalization mass $M_0$ is 
 defined as $M_0=8\times 10^{14}~h^{-1}~M_{\odot}$.}
 \label{fig:delc-m}
\end{figure}

\subsection{Mass function and number of clusters}
Galaxies and cluster of galaxies are embedded in the extended cold dark (CDM) matter haloes. 
In the Press-Schechter formalism, the abundance of CDM haloes in the Universe can be described in terms of their mass
and a Gaussian distribution function \citep{Press1974}. In fact, the fraction of the volume of the Universe which
collapses into an object of mass $M$ at a characteristic redshift $z$ is expressed by a Gaussian distribution function.
In the Press-Schechter formalism, the comoving number density of collapsed objects with masses in the range of $M$ and
$M+dM$ at the cosmic redshift $z$ can be written as \citep{Press1974,Bond1991}:
\begin{eqnarray}
 \frac{dn(M,z)}{dM}=\frac{\bar{\rho}_0}{M}\frac{d\nu(M,z)}{dM}f(\nu)\;,\label{eqn:PS1}
\end{eqnarray}
where $\bar{\rho}_0$ is the background density at the present time, and
\begin{equation}
 \nu(M,z)=\frac{\delta_{\rm c}}{\sigma}\;,
\end{equation}
where $\sigma$ is the r.m.s. of the mass fluctuations in spheres of mass $M$ and $f(\nu)$ is the mass function. 
The standard mass function presented by \cite{Press1974} differs from simulations at both high and low mass haloes 
\citep{Sheth1999,Sheth2002}. Therefore we will use the following mass function formula proposed by 
\cite{Sheth1999,Sheth2002}, the so called ST mass function
\begin{equation}
 f(\nu)= 0.2709\sqrt{\frac{2}{\pi}}\left(1+1.1096 \nu^{0.3}\right)
 \exp{\left(-\frac{0.707\nu^2}{2}\right)}\;.
 \label{eq:multiplicity_ST}
\end{equation}

The predictions of N-body simulations show that the abundance of clusters is well described by ST mass function up to 
$z\sim 2$ \citep{Jenkins:2000bv}. For higher redshifts, \cite{Jenkins:2000bv} showed deviations between the predictions 
of N-body simulations and the ST mass function. By using high-resolution N-body simulations, \cite{Reed:2003sq} 
predicted a fewer numbers of haloes than the ST mass function at $z\sim 15$. However, several other works showed that 
already beyond $z\gtrsim 2$, when using high resolution N-body simulations, the ST mass function is not a good fit 
\citep[see e.g.,][]{Klypin2011}. Here, nevertheless, this is not a big issue since our cluster sample spans a limited 
redshift range and in any case this is $z\lesssim 1$. 
To show how the results depends on the choice of the mass function, we compute the number of haloes in the redshift 
range $0\leqslant z\leqslant 1$ by using the prescription of \cite{Reed:2006rw} and the ST mass function. Our results 
show a difference less than $1\%$ for mass scales 
$1\times 10^{13}h^{-1}M\odot$, approximately $4\%$ for mass scales $5\times 10^{13}h^{-1}M\odot$ and roughly $11\%$ for 
mass scales $5\times 10^{14}h^{-1}M\odot$. For higher mass scales the difference may be as large as $20\%$ but these 
high mass clusters are very rare and can not affect our results. Notice that these results are in agreement with the 
results of \cite{Jenkins:2000bv}.

In a Gaussian density field, the amplitude of mass fluctuations $\sigma$ is given by
\begin{equation}\label{eq:sigma}
 \sigma^2(R,z)=\frac{1}{2\pi^2}\int_0^{\infty}{k^2P(k,z)}W^2(kR)dk\;,
\end{equation}
where $R$ is the comoving radius of the spherical overdense region, $W(kR)$ is the Fourier transform of a spherical 
top-hat filter and $P(k,z)$ is the linear matter power spectrum of density fluctuations at redshift $z$ 
\citep{Peebles1993}. The linear matter power spectrum at redshift $z$ can be written as
\begin{equation}\label{eq:power_matter}
 P(k,z)=P_{\rm 0}(k)T^2(k)D^2(z)\;,
\end{equation}
where we adopted the simple power law formula for $P_{\rm 0}$ as $P_{\rm 0}(k)=Ak^n_{\rm s}$ with the nearly
scale-invariant spectrum corresponding to $n_{\rm s}=0.96$. We also adopted the transfer function $T(k)$ which
considers baryonic features and was introduced in \cite{Eisenstein:1997ik}.

\begin{figure}
 \centering
 \includegraphics[width=0.45\textwidth]{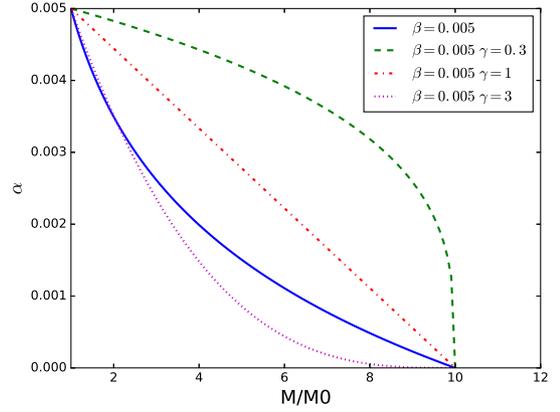}
 \caption{ Logarithmic and power-law parametrizations for $\alpha$ as a function of mass scale $M/M_{0}$. 
 The blue-solid, green-dashed, red-dotted-dashed and purple-dotted curves, represent the logarithmic, power-law with 
 $\gamma=0.3$, $\gamma=1$ and $\gamma=3$ parametrizations, respectively.}
 \label{fig:alpha_para}
\end{figure}

The comoving number density of clusters above a certain mass $M_0$ at collapse redshift $z$ is
\begin{equation}\label{eq:ndensity}
 n(>M_0,z)=\int_{M_0}^{\rm \infty}\frac{dn(M^{\prime},z)}{dM^{\prime}}dM^{\prime}\;.
\end{equation}
The comoving number of clusters per unit redshift with mass greater than a fiducial mass value $M_0$ is given by
\begin{equation}\label{eq:density1}
 N_{\rm bin}=n(M>M_0,z)\frac{dV}{dz}\;,
\end{equation}
where
\begin{equation}\label{eq:volune11}
 V(z)=4\pi\int_{0}^{z}\frac{d_{\rm L}^2(z^{\prime})}{(1+z^{\prime})^2H(z^{\prime})}dz^{\prime}\;,
\end{equation}
is the comoving volume at redshift $z$, and
\begin{equation}\label{eq:distance1}
 d_{\rm L}(z)=(1+z)\int_{0}^{z}\frac{dz^{\prime}}{H(z^{\prime})} \;,
\end{equation}
is the luminosity distance. It should be noted that the comoving volume depends on the cosmological model and hence 
volume effects will be introduced in the determination of the number counts. In the next section we will present the 
observational data on the abundance of massive galaxy clusters with masses above 
$M_0=8\times 10^{14}~h^{-1}~M_{\odot}$ within a comoving radius of $R_{\rm 0}=1.5~h^{-1}~Mpc$ at redshift 
$0.05\leq z\leq 0.83$. 
In this section we calculate the number of massive clusters $N_{\rm bin}$ with masses above $M_{\rm 0}$ as a function 
of redshift to investigate how shear and rotation can affect the number of clusters. We also compute the total 
number count of massive clusters above a given mass $M_0$ at the present time $z=0$ up to cosmic redshift $z$ as 
\begin{equation}\label{eq:density2}
 N=\int_{0}^{z}N_{\rm bin}dz^{\prime}=\int_{0}^{z}n(M>M_0,z^{\prime})\frac{dV}{dz^{\prime}}dz^{\prime}\;.
\end{equation}
Since there is not a theoretical expression describing the variation of $\alpha$ in terms of mass, we examine a 
logarithmic relation
\begin{equation}\label{eq:log-alpha}
 \alpha = -\beta \log_{10}\frac{M}{M_{\rm s}}\;,
\end{equation}
where $M_{\rm s}$ is a normalization mass and $\beta$ is a constant parameter representing the slope of the parameter 
$\alpha$. In what follows we set $M_{\rm s}=10M_0=8\times 10^{15}h^{-1}M_{\odot}$. Objects of such high mass are very 
rare, therefore we do not expect our analysis to be affected by this value. This is further justified by the fact that 
the abundance of massive cluster with $M>M_0$ is a factor of $10^{6}$ larger than clusters with $M>10M_0$. 
Notice that Eq.~(\ref{eq:log-alpha}) is a phenomenological relation and one can consider other possibilities such 
as a power-law parametrization
\begin{equation}\label{eq:alpha_power}
 \alpha = \beta \frac{\left(1-\frac{M}{M_{\rm s}}\right)^{\gamma}}{0.9^{\gamma}}\;,
\end{equation} 
 where the denominator is chosen to achieve the same value of $\alpha$ for the two different logarithmic and 
power-law parametrizations at $M_0$. In Fig (\ref{fig:alpha_para}) we show the evolution of $\alpha$ as a function of 
mass scale $M/M_{0}$ for both the logarithmic and the power-law parametrization. In all cases, we set the slope 
parameter as $\beta=0.005$ and examine different values of $\gamma$ for the power-law parametrization: $0.3$, $1.0$ and 
$3.0$. We see that both models have the same value $\alpha=0.005$ at mass scale $M=M_{0}$ and $\alpha=0$ for $M>10M_0$. 
Quantitatively speaking, in order to distinguish how the results may be changed for different parametrizations, we 
compute the number of massive clusters with mass scale $M>M_0$ in the redshift range $0<z<1$ using the ST mass 
function. We found that the differences between the logarithmic and the power-law parametrization are of the order of 
$0.1\%$, $0.5\%$ and $1.5\%$, respectively, for $\gamma=3$, $\gamma=1$ and $\gamma=0.3$. We see that such small 
differences would not affect our results. Since the logarithmic relation is defined with one parameter ($\beta$) to 
describe the mass dependency of $\alpha$, in what follows we adopt this parametrization and continue our analysis on 
the basis of Eq.~(\ref{eq:log-alpha}).
 
As mentioned above, the linear overdensity $\delta_{\rm c}$ is a crucial parameter in the Press-Schechter formalism and 
in Fig.~(\ref{fig:delc-m}) we show how the present time value of $\delta_{\rm c}$ varies with mass for three different 
values of $\beta$ with respect to the standard case with $\beta=0$. Increasing $\beta$ increases $\alpha$ which results 
into a higher value for the critical linear overdensity. In the case of $M=10M_{0}$, we see that for all models 
$\delta_{\rm c}$ tends to the fiducial value $1.675$ representing the standard $\Lambda$CDM value without shear and 
rotation. Note that in this case we set $\Omega_{\rm m}=0.3$.

\begin{figure*}
 \centering
 \includegraphics[width=0.45\textwidth]{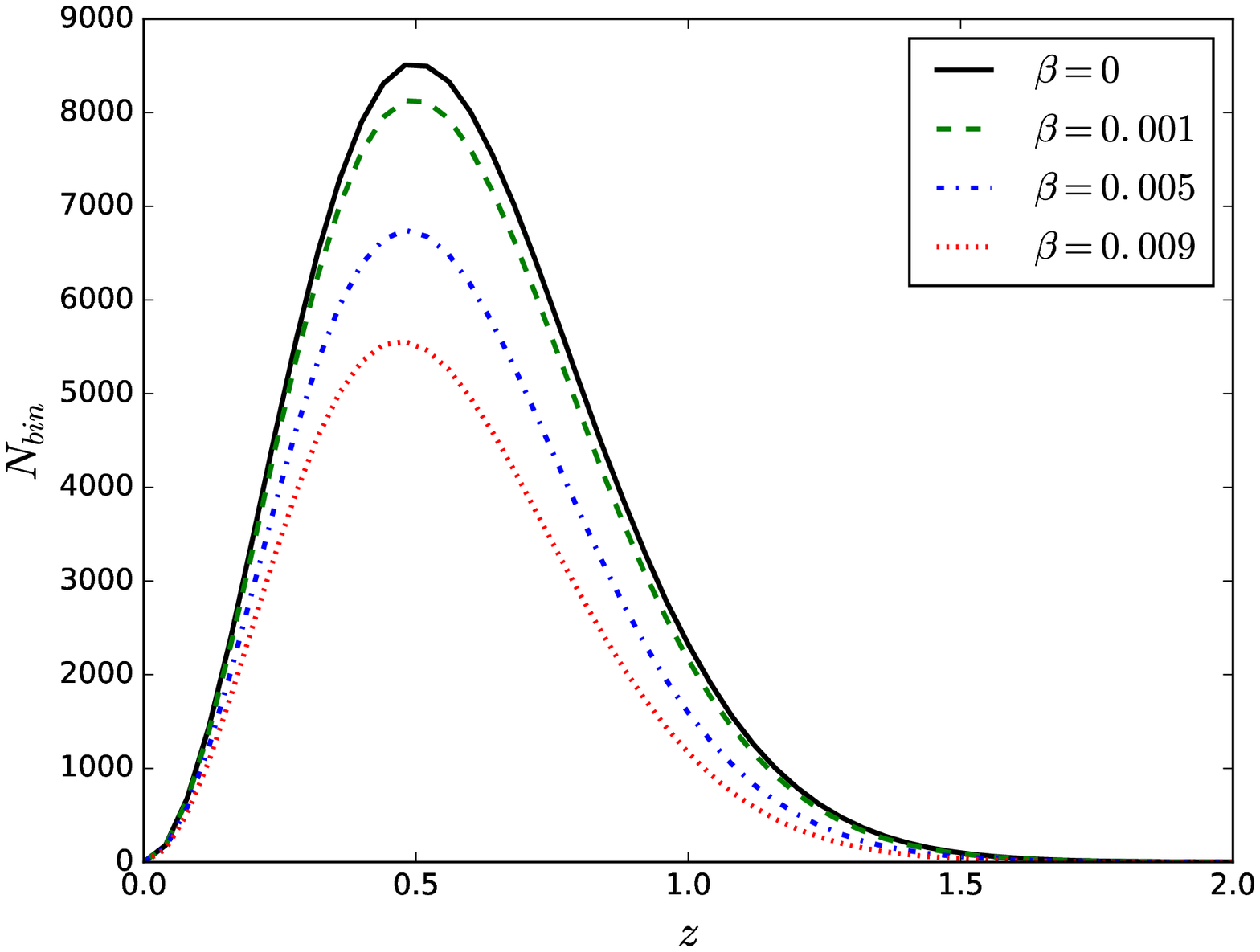}
 \includegraphics[width=0.45\textwidth]{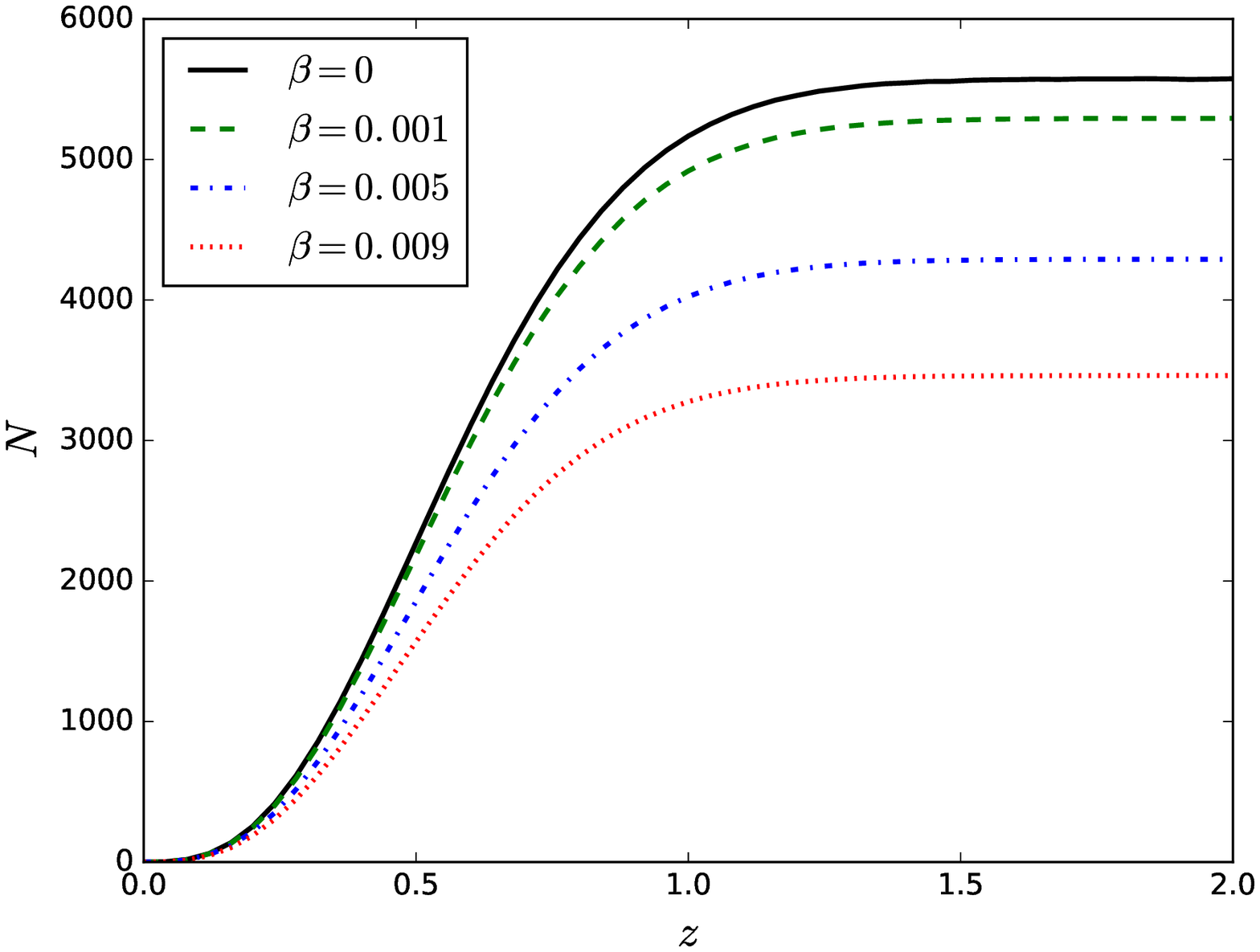}
 \caption{\textit{Left} (\textit{right}) \textit{panel}: Evolution of the number counts (total number counts) of 
 massive clusters as a function of the cosmic redshift for objects with masses above 
 $M_{\rm 0}=8\times 10^{14}h^{-1}M_{\odot}$ and for different values of the slope parameter $\beta$ computed in the 
 $\Lambda$CDM cosmology. Line styles and colours are as in Fig.~(\ref{fig:delc-m}).}
 \label{fig:number_co1}
\end{figure*}

In the left panel of Fig.~(\ref{fig:number_co1}) we show the number counts of massive clusters, $N_{\rm bin}$, above a 
given mass $M_0$ as a function of redshift $z$ for different values of the slope parameter $\beta$ as indicated in the 
legend. It can be seen that for higher values of $\beta$ we find less massive objects, in agreement with the general 
discussion above.

In the right panel of Fig.~(\ref{fig:number_co1}) we show the total number counts of massive clusters, $N$, in terms of 
the cosmic redshift for different values of the parameter $\beta$ as described in the legend. In analogy with what 
found for the number counts, we see that for higher values of $\beta$ massive galaxy clusters are less abundant. 
We also see that for $z\gtrsim 1$ we reach a plateau in the total number of clusters above a given mass $M_0$. 
This is due to the fact that such massive objects are not yet formed above $z\gtrsim 1$.

\section{Observational constraints on the \texorpdfstring{$\beta$}{b} parameter}\label{sect:4}
To constrain $\beta$ from observational data, we perform a likelihood analysis with current available observational 
data. After introducing the observational data used in this work, we will present and discuss the results. 
Since shear and rotation only affect the non-linear evolution of structures, to constrain $\beta$ we use the data 
related to number counts of massive galaxy clusters in different redshift bins as introduced in \cite{Campanelli2012}. 
Although a direct mass measurement is a difficult task, \cite{Campanelli2012} used a simple mass-temperature relation 
to estimate the number counts of massive galaxy clusters in four redshift bins with mass greater than 
$M_0=8\times10^{14}h^{-1}M_\odot$. We use the most recent geometrical probes like SNIa, cosmic microwave background 
(CMB) and baryon acoustic oscillation (BAO) to fix the geometry and add the data of massive galaxy clusters number 
count to find the best value of the slope $\beta$. Finally, since the relations in the non-linear regime we use to 
convert masses are well tested in simulations in $\Lambda$CDM models only, we restrict ourself to this cosmology.

We use the SnIa data from the Union 2.1 sample \citep{Union2.1:2012} which includes 580 SnIa over the redshift range 
$0<z<1.4$. The $\chi^2_{\rm sn}$ is
\begin{equation}\label{eq:xi2-sn}
 \chi^2_{\rm sn}=\mathbf{X}_{\rm sn}^{T}\mathbf{C}_{\rm sn}^{-1}\mathbf{X}_{\rm sn}\;,
\end{equation}
where $\mathbf{X}_{\rm sn}=\mu_{\rm th}-\mu_{\rm ob}$,
$\mu_{\rm th}(z)=5\log_{10}\left[(1+z)\int_0^z\frac{dx}{E(x)}\right]+\mu_0$ and $\mu_{\rm ob}$ is the observational
value of the distance module. For $\mathbf{C}_{\rm sn}^{-1}$ we use the covariance matrix including systematic 
uncertainties from \cite{Union2.1:2012}. 
Note that in this case the results are marginalized over the noisy parameter $\mu_0$ so that the result does not
depend on it.

For BAO, the $\chi^2_{\rm bao}$ is given by
\begin{equation}\label{eq:xi2-bao}
 \chi^2_{\rm bao}=\mathbf{Y}^{T}\mathbf{C}_{\rm bao}^{-1}\mathbf{Y}\;,
\end{equation}
where $\mathbf{Y}=(d(0.1)-d_{1},\frac{1}{d(0.35)}-\frac{1}{d_2},\frac{1}{d(0.57)}-
\frac{1}{d_3},d(0.44)-d_{4},d(0.6)-d_{5},d(0.73)-d_{6})$. The quantity $d(z)$ is defined through
\begin{equation}\label{eq:d(z)}
 d(z)=\frac{r_{\rm s}(z_{\rm drag})}{D_{\rm V}(z)}\;,
\end{equation}
where $r_{\rm s}(z_{\rm drag})$ is the comoving sound horizon at the drag epoch, namely the time when the baryons are 
"released" from the drag of the photons, and $D_{\rm V}(z)$ is a combination of the angular diameter distance and 
expansion rate of the Universe $H(z)$
\begin{equation}\label{eq:dv-bao}
 D_{\rm V}(z)=\left[(1+z)^{2}D^{2}_{\rm A}(z)\frac{z}{H(z)}\right]^{\frac{1}{3}}\;.
\end{equation}
We use the fitting formula for the redshift of the drag epoch, $z_{\rm drag}$, given in \cite{Eisenstein:1997ik} and
the covariance matrix $\mathbf{C}_{\rm bao}^{-1}$ introduced in \cite{Hinshaw:2012aka}. The observational data for 
$d_i$ are presented in Tab.~(\ref{tab:bao}).

\begin{table}
 \centering
 \caption{BAO data and their references.}
 \begin{tabular}{c|c|c|c}
 \hline
 \hline
 $z$      & $d_i$    & Survey   & References  \\ \hline
 $0.106$  & $0.336$  & 6dF      & \cite{Beutler:2011hx}\\ \hline
 $0.35$   & $0.113$  & SDSS-DR7 & \cite{Padmanabhan:2012hf}\\
 $0.57$   & $0.073$  & SDSS-DR9 & \cite{Anderson:2012sa} \\ \hline
 $0.44$   & $0.0916$ & WiggleZ  & \cite{Blake:2011en} \\
 $0.6$    & $0.0726$ & WiggleZ  & \cite{Blake:2011en} \\
 $0.73$   & $0.0592$ & WiggleZ  & \cite{Blake:2011en} \\ \hline \hline
 \label{tab:bao}
 \end{tabular}
\end{table}

CMB data can also be described through a covariance matrix $\mathbf{C}_{\rm cmb}^{-1}$,
\begin{equation}\label{eq:xi2-cmb}
 \chi^2_{\rm cmb}=\mathbf{X}_{\rm cmb}^{T}\mathbf{C}_{\rm cmb}^{-1}\mathbf{X}_{\rm cmb}\;,
\end{equation}
where
\begin{equation}\label{eq:x-cmb}
 \mathbf{X}_{\rm cmb}= \left(
 \begin{array}{c}
  R-R^{pl}\\
  l_{\rm a}-l_{\rm a}^{pl}\\
  \Omega_bh^2-(\Omega_bh^2)^{pl}
 \end{array}
 \right)\;,
\end{equation}
and the superscript "pl" refers to the Planck value and the quantities $l_{\rm a}$ and $R$ are defined as
$$ l_a = \pi\frac{D_{\rm A}(z_{\ast})}{r_s(z_{\ast})}\;, \quad
 R  =  \sqrt{\Omega_{\rm m}}H_{\rm 0}D_{\rm A}(z_{\ast})\;.$$
For $z_{\ast}$ (last scattering redshift), we use the fitting formula from \cite{Hu:1995en} and the covariance matrix 
from the analysis of \cite{Huang:2015vpa}.

To obtain the number counts of massive galaxy clusters in different redshift bins from a theoretical point of view, we 
follow the same procedure presented in \cite{Campanelli2012}. The redshift bins and the effective fraction of the 
observed comoving volume, $f_{\rm sky}$, are summarized in Tab.~(\ref{tab:redshift-bin}). 
 Data in bin 1 are extracted from a sample of 61 clusters introduced in \cite{Ikebe:2001xy}. The X-ray temperatures 
and fluxes of the cluster sample are measured with ASCA and ROSAT satellites. These clusters form a flux-limited 
complete sample in the redshift range $0<z<0.1$ and the effective fraction of observed comoving volume in this case is 
$0.309$. The data in bin $2$ are extracted from 14 massive clusters which are both redshift- and flux-limited 
\cite{Henry:2000bt}. The average redshift of these clusters is 0.38 and cover a fraction of 0.012 of the observed 
comoving volume. In \cite{Bahcall:1998ur,Bahcall:2002ru} the most massive clusters of galaxies at redshifts $z>0.5$ 
have been studied and used to constrain the cosmological parameters $\Omega_{\rm m}$ and $\sigma_8$. In bin $3$ we use 
the data of massive clusters in the redshift range $0.5<z<0.65$ with a tiny fraction of observed volume. To see the 
complete list of cluster data used here, we refer to \citep{Campanelli2012}. Note that since these data come from 
observational surveys which trace a specified portion of sky, we are not able to change these redshift bins in our 
analysis.

\begin{table*}
 \vspace*{-0.3cm}
 \caption{Redshift intervals for the number count data of massive galaxy clusters.
 $f_{\rm sky}(i)$ represents the effective fraction of the observed comoving volume of the $i^{\rm th}$ bin.}
 \begin{tabular}{ccccccc}
  \hline \hline
  & bin $i$  &$z_1^{(i)}$  &$z_2^{(i)}$  &$z_{\rm c}^{(i)}$  &Ref.  &$f_{\rm sky}(i)$ \\

  \hline

  &1  &0.00  &0.10 &0.050 &\cite{Ikebe:2001xy}    &0.309 \\
  &2  &0.30  &0.50 &0.375 &\cite{Henry:2000bt}    &0.012 \\
  &3  &0.50  &0.65 &0.550 &\cite{Bahcall:1998ur,Bahcall:2002ru} &0.006 \\
  &4  &0.65  &0.90 &0.825 &\cite{Donahue:1997sp} &0.001 \\

  \hline \hline

 \end{tabular}
 \label{tab:redshift-bin}
\end{table*}

Due to the low number of clusters in each bin, the Poisson statistics is used to define the likelihood function.
Therefore the $\chi^2_{\rm num}$, taking into account the uncertainty in the comoving number of clusters,
$\Delta N_{{\rm obs},i}$, is \cite[see][for details]{Campanelli2012}
\begin{eqnarray}\label{eq:chi2_num}
 \chi^2_{\rm num} = 2\sum_{i=1}^4 \left[ N_i - {N^{\prime}_{{\rm obs},i}}
 \left(1 + \ln N_i - \ln {N^{\prime}_{{\rm obs},i}} \right) \right] + \xi^2 \;,
\end{eqnarray}
where $N^{\prime}_{{\rm obs},i}=N_{{\rm obs},i}+\xi \Delta N_{{\rm obs},i}$ and $N_{{\rm obs},i}$ is the number of
clusters in each bin and is defined as
\begin{equation}\label{eq:number_obs}
 N_{{\rm obs},i}=f_{\rm sky}^{(i)}\int_{z_1^{(i)}}^{z_2^{(i)}}\int_{g(M_0)}^{\infty}
 \frac{dn(M^{\prime},z^{\prime})}{dM^{\prime}}dM^{\prime} \frac{dV}{dz^{\prime}}dz^{\prime}\;,
\end{equation}
where $g(M_0)$ relates the observed mass to the virial mass which depends on the model and the parameters. 
 Note that due to the asymmetric errors in X-ray temperature of clusters, the errors for the number of clusters in 
each bin $\Delta N_{{\rm obs},i}$ will also be asymmetric.
For the mass-concentration relation, we use the prescription of \cite{Prada:2011jf}. 
We use the procedure introduced in appendix B of \cite{Campanelli2012} to find $g(M_0)$. In 
equation~(\ref{eq:chi2_num}), $\xi$ is a uni-variate Gaussian random variable which is introduced to consider the 
uncertainty on the number of clusters in each bin. The values of $N_{{\rm obs},i}$ and $\Delta N_{{\rm obs},i}$ are 
shown in Tab.~{\ref{tab:data_num}}.

\begin{table*}
 \caption{Observational data for cluster number counts in different redshifts bins \citep{Campanelli2012}.}
 \begin{tabular}{llcccc}
  \hline \hline

  &   & bin 1  & bin 2  & bin 3  & bin 4 \\

  &   & $[T_{X,0}({\rm keV}) \,, N_{{\rm obs},1}]$  & $[T_{X,0}({\rm keV}) \,, N_{{\rm obs},2}]$
      & $[T_{X,0}({\rm keV}) \,, N_{{\rm obs},3}]$  & $[T_{X,0}({\rm keV}) \,, N_{{\rm obs},4}]$ \\

  \hline

  & $\Delta^{\prime}_{\rm vir} \in [25,175]$    & $[7.37\,, 5^{+1}_{-0}]$  & $[9.6\,, 0^{+0}_{-0}]$ &
    $[10.9\,, 0^{+1}_{-0}]$  & $[12.8\, ,0^{+1}_{-0}]$  \\

  & $\Delta^{\prime}_{\rm vir} \in ]175,375]$   & $[6.15\,, 15^{+2}_{-4}]$ & $[8.1\,, 1^{+0}_{-1}]$ &
    $[9.1\,, 1^{+1}_{-0}]$   & $[10.7\, ,1^{+0}_{-0}]$  \\

  & $\Delta^{\prime}_{\rm vir} \in ]375,750]$   & $[5.54\,, 21^{+2}_{-5}]$ & $[7.3\,, 1^{+4}_{-1}]$ &
    $[8.2\,, 2^{+0}_{-1}]$   & $[9.6\,, 1^{+0}_{-0}]$  \\

  & $\Delta^{\prime}_{\rm vir} \in ]750,1750]$  & $[5.14\,, 24^{+1}_{-1}]$ & $[6.7\,, 2^{+4}_{-1}]$ &
    $[7.6\,, 2^{+0}_{-1}]$   & $[8.9\,, 1^{+0}_{-0}]$  \\

  & $\Delta^{\prime}_{\rm vir} \in ]1750,3250]$ & $[4.91\,, 24^{+2}_{-0}]$ & $[6.4\,, 5^{+1}_{-3}]$ &
  $[7.3\,, 2^{+0}_{-0}]$     & $[8.5\,, 1^{+0}_{-0}]$  \\

  \hline \hline
 \end{tabular}
 \label{tab:data_num}
\end{table*}

Note that since the mass-temperature relation needs the virial overdensity with respect to the critical density, in the 
first column of Tab.~(\ref{tab:data_num}) we show the value of the virial overdensity $\Delta^{\prime}_{\rm vir}$
defined as
\begin{equation}\label{eq:virial_prime}
 \Delta^{\prime}_{\rm vir}=\frac{\Omega_{\rm m,0}(1+z)^3}{E^2(z)}\Delta_{\rm vir}\;,
\end{equation}
where $\Delta^{\prime}_{\rm vir}$ and $\Delta_{\rm vir}$ are the virial overdensity with respect to the critical and 
the background matter density, respectively. In the model considered here, $\Delta_{\rm vir}$ depends also on the mass 
as for the critical overdensity contrast. 
Since the virial overdensity in the range of mass $(M_0,10M_0)$ is practically constant  
\footnote{The virial overdensity varies about $0.2\%$ for the mass range between $(M_0,10M_0)$  and for the best 
fit parameters found in this work, its value is 268.6.},
we evaluate the virial overdensity for $M_0$ neglecting its mass evolution to compare with observations in 
Tab.~(\ref{tab:data_num}).

The overall likelihood function for all data sets used is
\begin{equation}\label{eq:like-tot}
 {\cal L}_{\rm tot}={\cal L}_{\rm sn} \times {\cal L}_{\rm bao} \times {\cal L}_{\rm cmb} \times {\cal L}_{\rm num}\;,
\end{equation}
which leads to
\begin{equation}\label{eq:like-tot_chi}
 \chi^2_{\rm tot}=\chi^2_{\rm sn}+\chi^2_{\rm bao}+\chi^2_{\rm cmb}+\chi^2_{num}\;.
\end{equation}

In our analysis, the free parameters are $P=\{\Omega_{\rm DM},\Omega_{\rm b},h,\beta,\xi\}$, where 
$\Omega_{\rm m} = \Omega_{\rm DM}+\Omega_{\rm b}$ is the total matter in the Universe. Other two relevant parameters 
($n_{\rm s}=0.9646$ and $\sigma_8=0.818$) are fixed to the Planck values \citep{Planck2015_XIII}. Our purpose in 
this analysis is to put a constraint on the slope of shear and rotation parameter $\beta$ appearing in the non-linear 
equations in the ESCM formalism by using the data for the number counts of massive galaxy clusters.

We perform a MCMC analysis to find the best fit values of the parameters as well as their uncertainties. In addition, 
our results are marginalized over $\xi$.  To do so, we use the GetDist package 
\footnote{\url{https://github.com/cmbant/getdist}} where for a large number of chains the package automatically 
marginalizes over a specific parameter. 
Results are summarized in Tab.~(\ref{tab:result}).

\begin{table}
 \caption{The best value of parameters and $1-\sigma$, $2-\sigma$ and $3-\sigma$ uncertainty intervals.}
 \label{tab:result}
 \begin{tabular}{|c| c c c c|}
  \hline
  \hline
  Parameters        & Best fit value          & $1-\sigma$   & $2-\sigma$           & $3-\sigma$ \\
  \hline
  $\Omega_{\rm m}$  & 0.284               & $\pm 0.0064$ & $^{+0.013}_{-0.012}$ & $^{+0.017}_{-0.016}$\\
  \hline
  $h$               & 0.678               & $\pm 0.017$  & $^{+0.032}_{-0.033}$ & $^{+0.038}_{-0.043}$\\
  \hline
  $\beta$          & $0.0019$ & $^{+0.0008}_{-0.0015}$   & $< 0.0043$        & $< 0.0054$ \\
  \hline
 \end{tabular}
\end{table}

\begin{figure}
 \centering
 \includegraphics[width=0.5\textwidth]{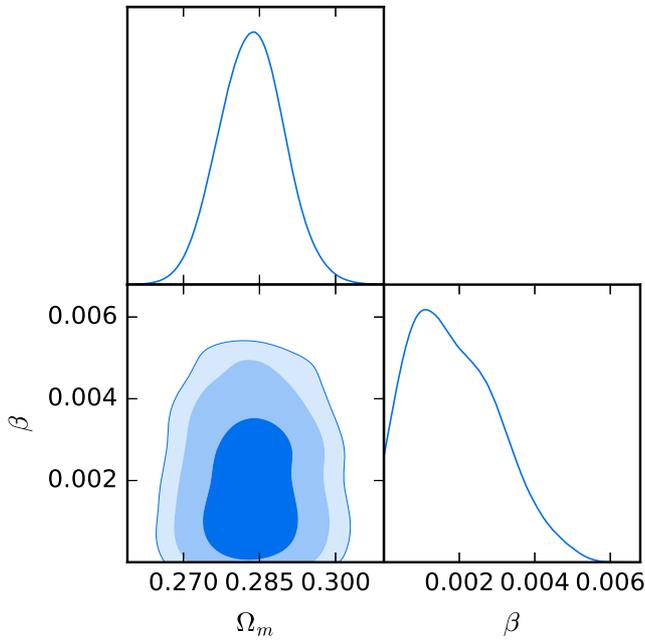}
 \caption{$1-\sigma$, $2-\sigma$ and $3-\sigma$ confidence regions for the ($\Omega_{\rm m},~\beta$) pair parameters.}
 \label{fig:omega_alpha}
\end{figure}

In Fig.~(\ref{fig:omega_alpha}) we show the confidence regions for the ($\Omega_{\rm m},~\beta$) pair parameters. In 
particular, while $\beta$ is constrained at the $1-\sigma$ confidence level, we find only an upper limit at the 
$2-\sigma$ and $3-\sigma$ confidence levels using massive clusters data. Since the posterior distribution for $\beta$ 
includes also the null value with a relatively high amplitude and data quality is relatively poor, we are only able to 
get an upper limit for the 2- and the 3-$\sigma$ confidence level. 
The best fit value of $\beta$ is $0.0019$ with a variance $^{+0.0008}_{-0.0015}$ at the $1-\sigma$ confidence level. 
Upper limits at the $2-\sigma$ and $3-\sigma$ confidence levels are shown in Tab.~(\ref{tab:result}). Getting tighter 
constraints requires more accurate data of massive clusters. 

 Few works, as mentioned in the Introduction, have studied the effect of the term $\sigma^2-\omega^2$ for the 
$\Lambda$CDM model, smooth and clustering dark energy models. More recently, \cite{Marciu2016} applied the same 
formalism to warm dark matter models (obtained with a non-null equation of state for the dark matter component) in 
different dark energy clustering models. The author assumed a constant value for the additional non-linear term 
$\beta=0.04$\footnote{In \cite{Marciu2016}, their $\beta$ corresponds to our $\alpha$.
which is valid for galactic scales. Note that in \cite{Marciu2016}, no scale dependence is assumed. 
Assuming our parametrization for $\alpha$ [Eq.~(\ref{eq:log-alpha})] and scales between $10^{11}~h^{-1}M_{\odot}$ and 
$10^{11}~h^{-1}M_{\odot}$, we find $\alpha\approx 0.01$ for $M\approx 10^{11}~h^{-1}M_{\odot}$ and 
$\alpha=7.4\times 10^{-3}$ for $M\approx 10^{12}~h^{-1}M_{\odot}$. This shows that the allowed value constrained 
by clusters would be a factor of four smaller than the one used by \cite{Marciu2016}. This implies a weaker effect on 
structure formation with respect to what found in the work.}

Our analysis allows us to compare the derived parameters $\Omega_{\rm m}$ and $h$ with recent work by the {\it Planck} 
team \citep{Planck2015_XIII}. Using distance prior from TT, TE, EE and low P data, we find $h=0.678\pm0.017$ at 
1-$\sigma$ level. This has to be compared with their value of $h=0.6727\pm0.0066$ showing a good agreement between the 
two determination. Slightly different is the comparison with the total matter parameter $\Omega_{\rm m}$. Our finding 
is $\Omega_{\rm m}=0.284\pm0.0064$ with respect to $\Omega_{\rm m}=0.3089\pm0.0062$, showing an approximately 
$2-\sigma$ tension with the Planck result. This is easily explained by taking into account that we keep the 
normalization of the matter power spectrum fixed and the evolution of the halo mass function is modified by the 
presence of the $\alpha$ term in the equations of motion for matter perturbations. As explained before, the effect of 
the parameter $\alpha$ is to decrease the amount of structures formed. This implies that to fit the resulting mass 
function we require a lower matter density parameter $\Omega_{\rm m}$ when $\sigma_8$ is held fixed. The Hubble 
parameter $h$ is largely unaffected by this, since we determine it by using distance prior data.

\begin{figure*}
 \centering
 \includegraphics[width=0.9\textwidth]{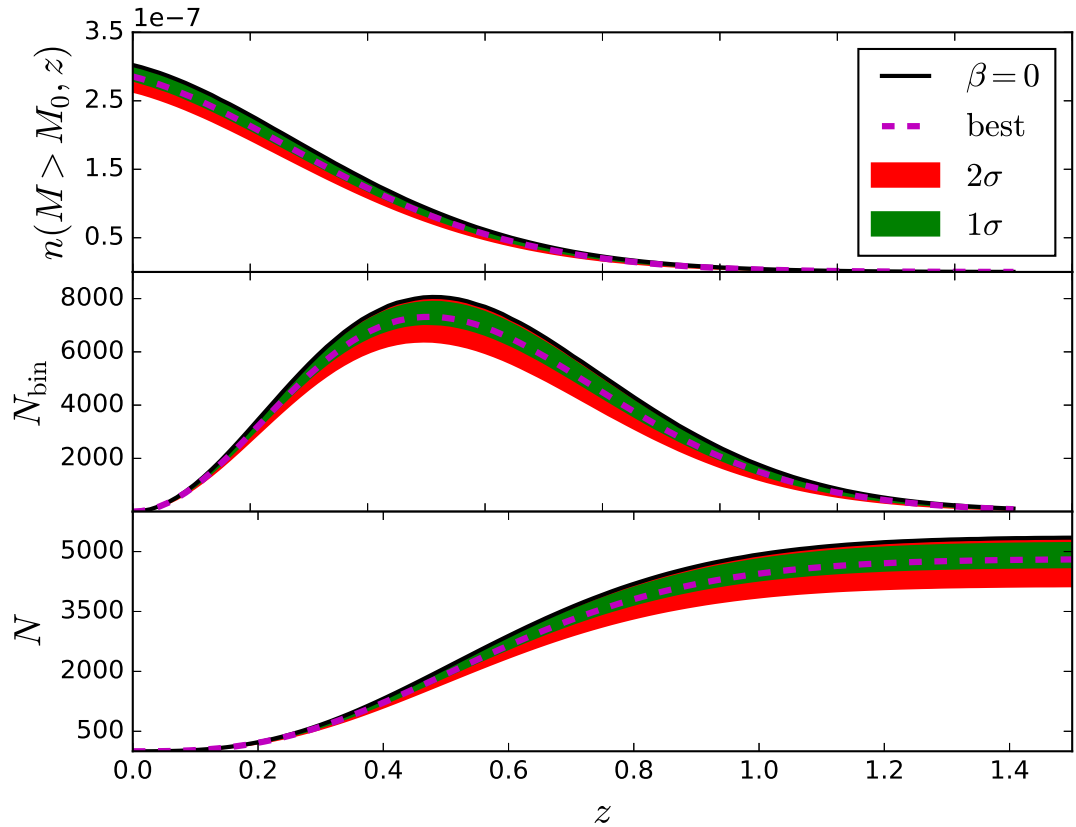}
 \caption{The computed comoving number density (top panel), the number of clusters in different redshift bins (middle 
 panel) and total number of massive clusters (bottom panel) for the best fit value of $\beta$ as well as the $1-\sigma$
 and the $2-\sigma$ confidence regions. 
 The magenta dashed line shows the best value and the black solid line is for the spherically symmetric case 
 ($\beta=0$).}
 \label{fig:mass_func_all}
\end{figure*}

Albeit small, a non-null value for $\beta$ will have an appreciable effect of the cumulative mass function, as shown in 
detail in Fig.~(\ref{fig:mass_func_all}), where we compare, as a function of cosmic redshift $z$ the comoving number 
density of objects above $M_0$ (top panel), the number of clusters in different redshift bins (middle panel) and the 
total number of clusters (bottom panel), respectively. We show results for the best fit value of $\beta$ together with 
the $1-\sigma$ and $2-\sigma$ bound regions. For the sake of comparison we also show the standard mass function and 
derived quantities for the case of spherical symmetry ($\beta=0$).

As expected, when $\beta\neq 0$, the number of cosmic objects formed is lower than in the standard case and the fact 
that differences are not negligible shows how much data need to improve, to have a reliable estimation of cosmological 
parameters based on cluster data. More quantitatively, the mass function at $z=0$, evaluated with the best fit values 
of $\beta$, is roughly $20\%$ smaller than that in the standard case with $\beta=0$. This decrement, of the order of 
the uncertainty on the mass function itself, translates to a lower number of objects in each redshift bin (middle 
panel). The maximum value for this quantity takes place approximately at $z\sim0.5$, showing that the peak of cluster 
abundance is reached at this epoch. It is interesting to see that this value is largely unaffected by the presence of 
shear and rotation terms. This is due to the fact that the parameter $\alpha$ only slows down structure formation, but 
it does not affect its physics. Differences are in this case more limited and are of the order of $9\%$ around the 
peak. Another indication that only the total number of objects is affected comes from the bottom panel, where we 
present the total number of objects. Note that the distribution flattens, reaching a plateau, around $z\gtrsim1$, since 
at higher redshifts massive objects are not formed yet. Differences between the spherically symmetric case and the 
extended one are about $8\%$.

It is also interesting to compare our results with similar works in literature. Remember though that in our analysis we 
do not allow the normalization of the matter power spectrum to change.\\
\cite{Campanelli2012}, using only cluster data, found $\Omega_{\rm m}\geq 0.38$ and at the same time 
$\sigma_8\leq 0.69$ at the $1-\sigma$ confidence level. Both results are in tension with our determination. This can be 
explained by the fact that we used cluster data with recent background data to constrain our model and the geometry 
being fixed in our analyses. Note that results of \cite{Campanelli2012}, by considering the background data, are in 
well agreement with our results. On the other hand \cite{Mantz2010}, using a constant dark energy model, find 
$\Omega_m=0.23\pm 0.04$ and $\sigma_8=0.82\pm 0.05$ which are compatible with our results. In addition, our results are 
in complete agreement already at the $1-\sigma$ confidence level with \cite{2009ApJ...692.1060V} and 
\cite{Schuecker2003} which are in good agreement with results of \cite{Campanelli2012}. Finally, it is worth to mention 
that these same data were used also by \cite{Bahcall:1998ur} and \cite{Bahcall:2002ru} who found low values for the 
matter density parameter ($\Omega_{\rm m}=0.17\pm0.05$) and high normalization ($\sigma_8=0.98\pm0.10$). These results 
are both in disagreement with the ones found by \cite{Campanelli2012} and in this work. This shows how a different 
treatment of the data and different data set will lead to different results.

\section{Conclusion}\label{sect:5}
Massive galaxy clusters, being at the high mass end of the mass function are becoming a common tool in cosmology. Their 
abundance is a strong indicator of non-linear structure formation and it depends on the value of important cosmological 
parameters, such as the matter density parameter $\Omega_{\rm m}$, the mater power spectrum normalization $\sigma_8$ 
and the dark energy equation of state $w_{\rm de}$. A precise determination of the mass function is a current goal of 
both theoretical and observational studies, due to the wealth of implications related to it.

From a theoretical point of view, the mass function is related to the function $\delta_{\rm c}$, that, in the framework 
of the spherical collapse model, represents the density above which structures can form. In the standard approach, 
perturbations are assumed to be spherical and non rotating, but in an era of precision cosmology it is necessary to 
relax this assumption. Shear and rotation can be added naturally into this formalism as shown recently by 
\cite{Popolo2013a,Popolo2013b,Pace2014a} and their combination is parametrized via the parameter $\alpha$. This 
extension of the simple spherically symmetric model makes such that $\delta_{\rm c}$ is now a function of both mass and 
redshift, contrary to the standard case where it only depends on time. This implies that the mass function and hence 
the total number of objects that can be observed will strongly depend on the evolution with mass of the parameter 
$\alpha$. Since theory, so far, does not constrain it, in this work we choose a particularly simple form: 
$\alpha=-\beta\log_{10}\frac{M}{M_s}$, where $\beta$ is the slope of the logarithmic relation and 
$M_s=8\times 10^{15}~h^{-1}~M_{\odot}$ is a normalization mass. When $M=M_s$, deviation from sphericity are null and we 
recover the standard case.

The combined effect of shear and rotation, due to the dominance of the latter, implies a decreased number of objects 
with respect to the spherically symmetric case since structure formation is slowed down.

Using data on massive clusters by \cite{Campanelli2012} we constrain, for the first time to our knowledge, the 
value of the slope $\beta$ and we infer its consequences on the number of massive objects. In our analysis we find 
$\Omega_{\rm m}=0.284\pm0.0064$, $h=0.678\pm0.017$ and $\beta=0.0019^{+0.0008}_{-0.0015}$ at $1-\sigma$ level, when 
keeping $\sigma_8=0.818$ fixed and restricting our analysis to a flat $\Lambda$CDM model. The value for $h$ is in 
complete agreement with \cite{Planck2015_XIII} but we find a slight tension for the value of $\Omega_{\rm m}$. 
This is due to the fact that, when fixing the normalization of the matter power spectrum, a decrement in the mass 
function requires a lower $\Omega_{\rm m}$. This has as consequence a decrement of about $9\%$ in the number of massive 
clusters.

Our result for $\Omega_{\rm m}$ is in agreement with results in \cite{Mantz2010,Schuecker2003,2009ApJ...692.1060V} 
which use massive clusters to constrain cosmological parameters. In addition our results are compatible with results of 
\cite{Campanelli2012} when they combine geometrical data to the massive clusters data.
 At the same time, using the same data, \cite{Campanelli2012} is in disagreement with \cite{Bahcall:1998ur} and 
\cite{Bahcall:2002ru} who found a very low (high) value for $\Omega_{\rm m}$ ($\sigma_8$). This shows how results can 
be dramatically different when performing a different analysis and the importance of having good quality data.

We conclude therefore that despite the data have room for $\beta$ at the order of per mill, it is necessary to have 
better data to constrain this value better, since as shown in Fig.~\ref{fig:omega_alpha}, we are only able to give 
upper limits for it at the 2- and $3-\sigma$ level.

\section*{Acknowledgements}
The authors thank the anonimous referee whose comments helped to improve the scientific content of this work. 
FP is supported by an STFC postdoctoral fellowship.

\bibliographystyle{mnras}
\bibliography{ref}

\label{lastpage}

\end{document}